\documentclass{osa-article}

\journal{osajournal}
\usepackage{lineno}
\usepackage{bm}
\usepackage{adjustbox}
\usepackage{multirow}
\articletype{article}
\begin{document}

\title{Ultracompact 4H-silicon carbide optomechanical resonator with $f_m\cdot Q_m$ exceeding $10^{13}$ Hz}
\author{yuncong Liu \authormark{1}, Wenhan Sun \authormark{2}, Hamed Abiri \authormark{3}, Philip X.-L. Feng \authormark{1,*}, and Qing Li\authormark{2,*}}

\address{\authormark{1}Department of Electrical and Computer Engineering, University of Florida, Gainesville, FL 32611, USA\\
\authormark{2}Department of Electrical and Computer Engineering, Carnegie Mellon University, Pittsburgh, PA 15213, USA\\ 
\authormark{3} School of Electrical and Computer Engineering, Georgia Institute of Technology, Atlanta, GA 30332 USA\\}
% \authormark{3}Currently with the Department of Electronic Journals, Optica Publishing Group, 2010 Massachusetts Avenue NW, Washington, DC 20036, USA}

\email{\authormark{1,*}philip.feng@ufl.edu} %% email address is required; see note below about the corresponding author designation
\email{\authormark{2,*}qingli2@andrew.cmu.edu}

\begin{abstract}
Silicon carbide (SiC) has great potential for optomechanical applications due to its outstanding optical and mechanical properties. However, challenges associated with SiC nanofabrication have constrained its adoption in optomechanical devices, as embodied by the considerable optical loss or lack of integrated optical access in existing mechanical resonators. In this work, we overcome such challenges and demonstrate a low-loss, ultracompact optomechanical resonator in an integrated 4H-SiC-on-insulator (4H-SiCOI) photonic platform for the first time. Based on a suspended $4.3$-$\mu$m-radius microdisk, the SiC optomechanical resonator features low optical loss ($<1$ dB/cm), a high mechanical frequency $f_m$ of $0.95 \times 10^9$ Hz, a mechanical quality factor $Q_m$ of $1.92\times10^4$, and a footprint of $<1\times 10^{-5}$ mm$^2$. The corresponding $f_m\cdot Q_m$ product is estimated to be $1.82 \times 10^{13}$ Hz, which is among the highest reported values of optomechanical cavities tested in an ambient environment at room temperature. In addition, the strong optomechanical coupling in the SiC microdisk enables coherent regenerative optomechanical oscillations at a threshold optical dropped power of 14 $\mu$W, which also supports efficient harmonic generation at increased power levels. With such competitive performance, we envision a range of chip-scale optomechanical applications to be enabled by the low-loss 4H-SiCOI platform.   
\end{abstract}

\section{Introduction}
Advances in nanofabrication and photonic engineering have propelled cavity optomechanical systems to a pivotal role for probing fundamental optical force-mediated mechanical phenomena and enabling precision control over light-matter interactions \cite{aspelmeyerRMP2014cavity}. These optomechanical systems have evolved rapidly with improved coupling efficiencies and tunable responses, promising innovations in vast fields including metrology \cite{hossein_PRA2006_char,tallur_OptExpress_2011_monolithic,luan_SciRep_2014_integrated}, quantum information\cite{safavi_Nature_2013_squeezed,bochmann_NatPhys_2013_nanomechanical,balram_NatPhotonics_2016_coherent}, and advanced sensing\cite{wu_NatNanotechnol_2017_nanocavity,li_Nanophotonics_2021_cavity}. Despite the significant progress achieved with conventional materials such as silicon (Si)\cite{sun_APL_2012_high,schwab_MN_2022_very,sonar_Optica_2025_high, wu_SciAdvances_2023_chip}, silicon dioxide (SiO$_2$)\cite{kippenberg_PRL_2005_analysis,jiang_OptExpress_2015_chip,wang_PRX_2024_taming}, silicon nitride ($\text{Si}_3\text{N}_4$)\cite{liu_PRL_2013_electromagnetically,grutter_APLPhotonics_2018_invited}, the quest for better performance has driven attention toward other candidates such as aluminum nitride (AlN)\cite{han_APL_2015_10-ghz,sohn_NatPhotonics_2018_time}, diamond\cite{Mitchell_Optica_16_diamond,burek_Optica_2016_diamond}, gallium phosphide (GaP)\cite{schneider_Optica_2019_optomechanics,chen_NC_2023_optomechanical}, and lithium niobate (LN)\cite{jiang_SciRep_2016_chip,shen_APL_2020_high,jiang_Optica_2019_lithium} for their broadband optical transparency, strong optical confinement, and exceptional mechanical properties that enable high-frequency mechanical modes with minimal dissipation. Among them, silicon carbide (SiC) stands out due to its exceptional thermal stability, mechanical robustness, and compatibility with CMOS processing \cite{falk_NC_2013_polytype,lu_APL_2020_silicon,sementilli_NanoLetters_2025_low,liu_CommuniEngi_2024_4h}. For example, the mechanical frequency and quality factor product ($f_m \cdot Q_m$) of SiC is predicted to be $6.4\times 10^{14}$ Hz in the Akhiezer regime, which is one order of magnitude higher than that of Si ($3.9 \times 10^{13}$ Hz) and diamond ($3.7 \times 10^{13}$ Hz)\cite{chandorkar2008limits,ayazi2011energy,ghaffari_SciRep_2013_quantum}. However, the very characteristics that make SiC attractive — its high mechanical rigidity and chemical inertness — also present fabrication challenges that constrain the realization of high-performance SiC optomechanical resonators. To date, only a few experiments have successfully demonstrated optomechanical resonators in the polytype of 3C-SiC\cite{lu_SciRep_2015_high}. In 4H-SiC, the reported SiC mechanical resonators are all bulky and only support mechanical modes with relatively low frequencies ($<10$ MHz) \cite{sementilli_NanoLetters_2025_low,liu_CommuniEngi_2024_4h}. In addition, the lack of integrated optical access necessitates additional excitation mechanisms, further complicating device design and scaling potential.

\begin{figure}[htbp]
\centering
\includegraphics[width=0.85\linewidth]{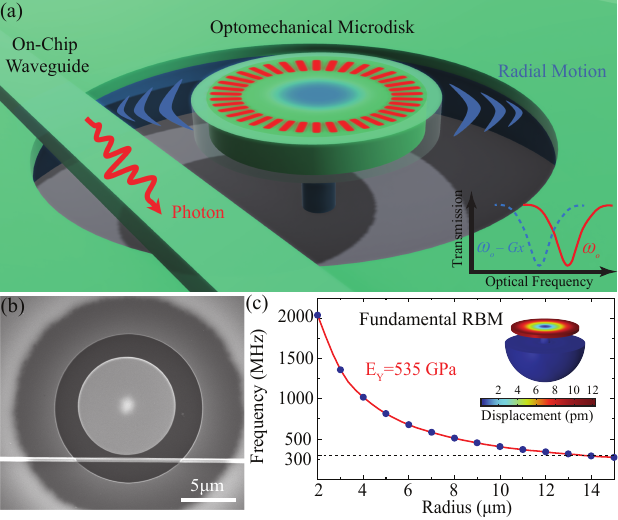} 
\caption{(a) Schematic of the silicon carbide (SiC) cavity optomechanical system and its operating principle. (b) Scanning electron micrograph of a $4.5$-$\mu$m-radius microdisk with on-chip waveguide access. The 4H-SiC microdisk has a device thickness of 600 nm, sitting on a 2-$\mu$m-thick silicon dioxide pedestal with an approximate undercut width of $3.4\ \mu$m. (c) Simulated mechanical frequency of the fundamental radial breathing mode (RBM) as a function of the radius. The inset displays the corresponding mechanical displacement profile based on finite element simulation:Young’s modulus $E_\text{Y}$ = $535$ GPa, Poisson's ratio $\nu=0.183$, and mass density $\rho =$ 3210 $\text{kg/m}^3$).}
\label{Fig_schematic}
\end{figure}

In this work, we report the first demonstration of a high-performance, integrated 4H-SiC optomechanical resonator that simultaneously functions as a high-finesse optical cavity. As illustrated in Fig.~\ref{Fig_schematic}, our design employs an ultracompact (radius around $4$ $\mu$m) microdisk resonator suspended on a low-loss 4H-SiC-on-insulator (4H-SiCOI) platform, which supports whispering gallery modes in telecom wavelengths with intrinsic optical quality factors up to $\sim1 \times10^6$. The strong optomechanical coupling resulting from highly co-localized optical and mechanical modes enables sensitive readout of the Brownian motion through optical photodetection, achieving a displacement sensitivity of ~0.144 $\text{fm}/\sqrt{\text{Hz}}$. With this method, we characterize the fundamental mechanical mode to exhibit a mechanical frequency ($f_m$) of 950 MHz and a mechanical quality factor ($Q_m$) near $19,200$. The corresponding mechanical frequency and quality factor product ($f_m \cdot Q_m$) is estimated to be $1.82\times 10^{13}$ Hz, which is among the highest reported metrics of optomechanical cavities tested in an ambient environment at room temperature. Finally, we also achieve radiation pressure-driven regenerative (or self-sustained) optomechanical oscillations and harmonic generation with an on-chip optical power well below 1 mW. 

\section{Device design and fabrication}
An optomechanical resonator requires strong coupling between the optical and mechanical modes. As illustrated in Fig.~\ref{Fig_schematic}, the optical whispering-gallery mode supported by the SiC microdisk generates radiation pressure along the radial direction, efficiently driving the fundamental mechanical radial-breathing mode with dominant radial displacement (see Fig.~\ref{Fig_schematic}a). For such mechanical modes, numerical simulations based on the finite element method (FEM) point to a mechanical frequency above the very-high-frequency (VHF) band (i.e., $>300$ MHz) for SiC microdisks with radii below $13\ \mu$m (see Fig.~\ref{Fig_schematic}c). In addition, the optomechanical coupling coefficient resulting from the moving-boundary effect scales as $g_{\text{om}}\approx  \omega_\text{o}/R$, where $\omega_\text{o}$ is the angular optical resonance frequency and $R$ is the radius. While a small $R$ benefits a high mechanical frequency and strong optomechanical coupling, the optical quality factor of the SiC microdisk could drop significantly if the radius is close to the optical wavelength ($1.55\ \mu$m in this experiment). As such, we choose to work with disks with radii near $4\ \mu$m to balance the mechanical and optical properties. A detailed simulation shows that a 4-$\mu$m-radius SiC disk exhibits a mechanical frequency $f_m \approx 1$ GHz, an effective mass of $m_{\text{eff}} \approx 71$ pg, and $g_{\text{om}}/2\pi \approx 44$ GHz/nm for the fundamental transverse-electric (TE$_{00}$) mode. These numbers translate to a vacuum optomechanical coupling rate $g_0 = g_{\text{om}}/\sqrt{\hbar/(4\pi m_{\text{eff}}f_m)} \approx 15$ kHz ($\hbar\equiv h/2\pi$ with $h$ being the Planck's constant) for the interaction between optical and mechanical modes. 

Proceeding to nanofabrication, a 4H-SiCOI chip consisting of 700-nm-thick 4H-SiC layer on top of 2-$\mu$m-thick silicon dioxide (NGK Insulators) is employed for SiC disks with radii varied from $4.3\ \mu$m to $4.7\ \mu$m \cite{Li_4HSiC_comb}. The device fabrication begins with defining photonic structures using a negative-tone resist (flowable oxide 16) in  ebeam lithography (EBL), which is followed by fluorine-based plasma dry etching to target 600 nm removal of SiC. Next, a positive-tone ebeam resist (PMMA) is employed to pattern circular release windows surrounding the microdisk resonators, within which the remaining 100 nm SiC pedestal is etched away for the subsequent undercutting process (see Fig.~\ref{Fig_schematic}b). Note that during this step, the top surface of the SiC microdisks is no longer protected by resist, resulting in an approximate 100 nm reduction in the SiC thickness (i.e., the final microdisk thickness is around 600 nm instead of the original 700 nm). After the two-step EBL and dry etching process, the SiC chip is finally dipped in buffered oxide etch solution to isotropically remove the silicon dioxide layer underneath the SiC microdisk. The wet etch time is carefully controlled to maximize the undercutting ratio while avoiding structural collapse in microdisk resonators. In practice, we manage to attain an undercut ratio (undercut width normalized by the disk radius) close to 80$\%$ for $4.3$-$\mu$m-radius SiC microdisks. For larger disks with radii including $4.5$ $\mu$m and $4.7$ $\mu$m, the undercut width is similar but the undercut ratio becomes smaller.

\begin{figure}[htbp]
\centering
\includegraphics[width=0.9\linewidth]{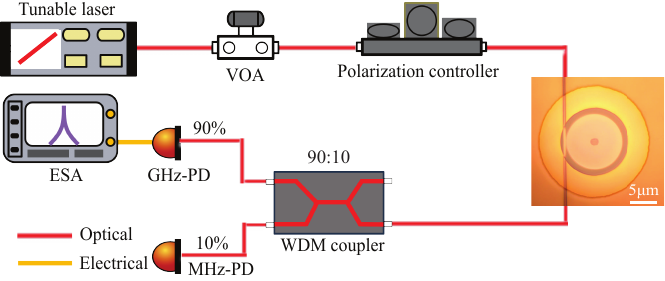} 
\caption{Experimental setup for the optomechanical measurement, where the slow photodetector (MHz-PD) is to identify optical resonances and the fast photodetector (GHz-PD) is for the mechanical characterization. VOA: variable optical attenuator; WDM: wavelength division multiplexer; PD: photodetector; and ESA: electrical spectrum analyzer.}
\label{Fig_setup}
\end{figure}

\section{Results}
The experimental setup for the optomechanical characterization of the SiC chip is shown in Fig.~\ref{Fig_setup}, which is carried out at room temperature and atmosphere. Briefly, we first perform linear transmission scans at low optical powers to identify the resonant modes of SiC microdisks, focusing on the TE$_{00}$ mode family which is expected to exhibit the highest optical quality factors in the telecom band. Next, we fix the laser detuning relative to the optical resonance and characterize the mechanical properties and optomechanical interactions by sending the detected signal to an electrical spectrum analyzer (ESA). For these purposes, a narrow-linewidth tunable laser (linewidth < 100 kHz with a tuning range of 1510-1640 nm) is employed, whose output power is fixed at 5 mW and attenuated externally using an in-loop variable optical attenuator (0-60 dB). In addition, the polarization state of light is manually adjusted with the help of a fiber polarization controller. To achieve efficient coupling between the fiber and the SiC chip, a pair of grating couplers are designed for each microdisk in the 1550 nm band, attaining a total insertion loss of 10-13 dB \cite{Li_4HSiC_comb}. After transmission, the collected signal is split into two paths using a 10:90 fiber coupler: the 10$\%$ portion is connected to a low-speed (MHz) photodetector (Thorlabs PDB450C) with large electronic gains for the resonance scan, while the 90$\%$ portion is sent to a high-speed photodetector (Newport AD-40 with 12 GHz bandwidth) followed by a real-time ESA (Tektronix RSA5106 with $6.2$ GHz bandwidth).

\subsection{Optical characterization}
\begin{figure}[htbp]
\centering
\includegraphics[width=0.9\linewidth]{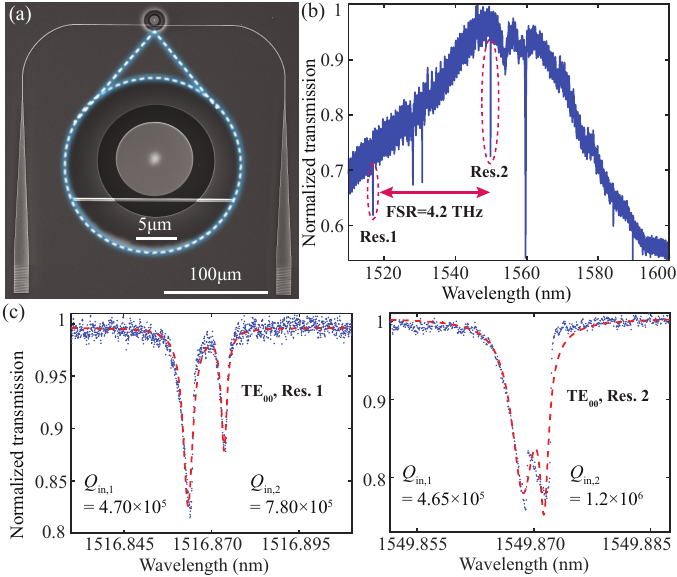} 
\caption{(a) Scanning electron micrograph of a waveguide-coupled SiC microdisk. The inset shows a close-up view of the suspended microdisk. (b) Representative transmission scan of a suspended $4.5$-$\mu$m-radius microdisk, with two adjacent azimuthal orders of the TE$_{00}$ mode family identified with a free spectral range (FSR) of $4.2$ THz. (c) Zoom-in plots of the two TE$_{00}$ resonances highlighted in (b), both of which exhibit mode splitting. The red dashed line represents numerical fitting using a doublet model, revealing intrinsic quality factors in the range of ($0.5$-$1.2$) $\times 10^6$.}.
\label{Fig_highQ}
\end{figure}

Systematic linear optical characterization is performed for SiC microdisks with radii ranging from $4.3\ \mu$m to $4.7\ \mu$m. The example provided in Fig.~\ref{Fig_highQ} corresponds to a waveguide-coupled $4.5$-$\mu$m-radius suspended SiC microdisk. As shown by the scanning electron micrograph in Fig.~\ref{Fig_highQ}a, the access waveguide in the coupling region is tapered down to 600 nm in width to increase the field overlap with the confined resonant modes. The linear transmission scan displayed in Fig.~\ref{Fig_highQ}b confirms that the TE$_{00}$ mode family is efficiently excited, which has an estimated free spectral range of $4.2$ THz. The zoom-in plots of two TE$_{00}$ resonances highlighted in Fig.~\ref{Fig_highQ}b reveal mode splitting resulting from the coupling between the clockwise and counterclockwise modes induced by the sidewall roughness \cite{YiXu_mode_splitting}. Furthermore, numerical fitting based on a doublet model points to an intrinsic optical quality factor up to $1.2$ million. It is worth noting that the intrinsic quality factors of the TE$_{00}$ mode family exhibit variations among different azimuthal orders and devices, with the majority falling in the range of $0.3$-$0.8$ million. Such variations are consistent with under-coupled microresonators with their loss dominated by the scattering from the roughness introduced to sidewalls in nanofabrication \cite{Li_azimuth}. 

\subsection{Mechanical mode characterization}
\begin{figure}[htbp]
\centering
\includegraphics[width=0.9\linewidth]{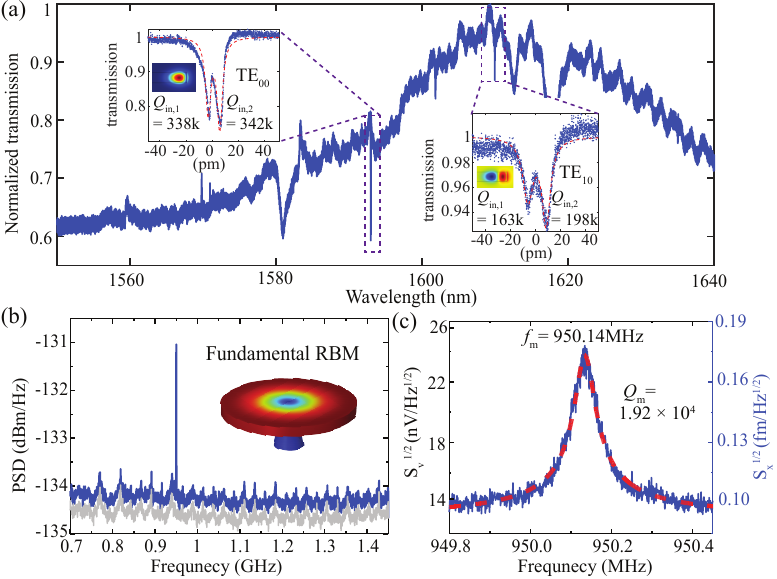} 
\caption{(a) Linear transmission of a suspended $4.3$-$\mu$m-radius SiC microdisk resonator. The insets are the zoomed resonances for the TE$_{00}$ and TE$_{10}$ modes at 1592 nm and 1610 nm, respectively, with the red dashed lines representing numerical fitting based on a doublet model.(b) Optically transduced electrical spectrum (blue curve) of the fundamental radial breathing mode (RBM) measured with an approximate dropped power of $\sim$2.5 $\mu$W. The gray trace corresponds to the background noise level of the high-speed photodetector (i.e., no optical input). The resolution bandwidth of the ESA is set at 500 Hz. (c) Close-up view of the fundamental RBM around 950 MHz with a damping-limited quality factor of 19,200 at room temperature (resolution bandwidth of ESA set at 20 Hz). The left and right y-axes correspond to the measured data in (b) converted to the voltage and displacement domain, respectively. The red dashed line is numerical fitting based on a damped simple harmonic resonator model.}
\label{Fig_R4p3}
\end{figure}

Next, we proceed to the characterization of the mechanical properties of SiC microdisks. In this regard, the undercut ratio, defined as the undercut width normalized by the radius of the microdisk, is a critical parameter determining the intrinsic mechanical loss. Given that the undercut width is similar among microdisk resonators with different radii (varied from $4.3$ $\mu$m to $4.7$ $\mu$m), the $4.3$-$\mu$m-radius SiC microdisk is expected to exhibit the highest mechanical $Q$s due to its largest undercut ratio. Figure \ref{Fig_R4p3}a shows the optical resonance measurement for a $4.3$-$\mu$m-radius SiC microdisk, where the TE$_{00}$ mode is found to possess an intrinsic optical quality factor of  $3.4\ \times10^5$ at 1592 nm, along with a higher-order TE$_{10}$ mode with an intrinsic quality factor near $1.8\ \times 10^5$ at 1610 nm.

The large optical quality of the observed WGMs combined with strong optomechanical coupling enables efficient optical transduction of mechanical motion. To measure the mechanical resonance, we fix the laser frequency to the blue side of the optical resonance while recording the optical transmission using a high-speed photodetector and an electrical spectrum analyzer (ESA). The optical power dropped into the microdisk ($P_d$) from the coupling waveguide is defined as $P_d=P_{in}(1-T_o)$, with $P_{in}$ and $T_o$ representing the input optical power in the waveguide and the normalized optical transmission, respectively. The optical dropped power is initially maintained at a low level ($\approx 2.5\ \mu$W) to minimize dynamic optical back-action, so that the thermomechanical (Brownian) motion of the microdisk dominates in the detected signal. As shown in Fig.~\ref{Fig_R4p3}b, the fundamental RBM at $950.1$ MHz is clearly observed in the ESA spectrum, which agrees with the simulated value $945.93$ MHz in Fig.~1c reasonably well. The difference is mainly attributed to the uncertainty in estimating the undercut width and disk radius. Moreover, we can extract the mechanical quality factor ($Q_m$) by fitting the measured RF spectrum to a damped simple harmonic resonator model\cite{jia_JMEMS_2019_very}, revealing a $Q_m$ of $1.92 \times 10^4$ (see Fig.~\ref{Fig_R4p3}c). Note that this value does not vary with the optical dropped  power provided that it is small enough ($P_d<10\ \mu$W), suggesting that the mechanical resonator is mainly limited by damping effects with negligible contribution from the optical back-action. Consequently, the fundamental RBM mode achieves a $f_m\cdot Q_m$ product of $1.82 \times 10^{13}$ Hz, which is on par with the highest values achieved among the reported WGM-type optomechanical microresonators (a more detailed comparison is provided in Table 1)\cite{han_APL_2015_10-ghz,Mitchell_Optica_16_diamond}.

\begin{figure}[htbp]
\centering
\includegraphics[width=0.65\linewidth]{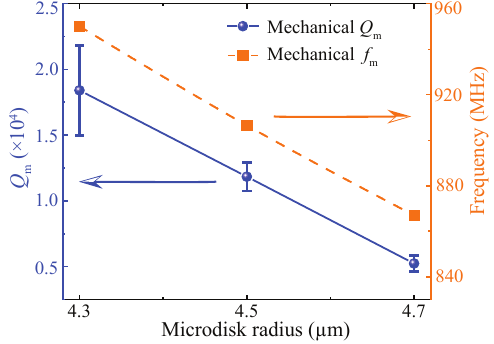} 
\caption{Summary of measured mechanical quality factors ($Q_m$, left axis) and frequencies ($f_m$, right axis) of SiC microdisks with different radii. The degradation of the mechanical $Q_m$ with increased radius is mainly attributed to the reduced undercut ratio.}
\label{Fig_all}
\end{figure}

The displacement sensitivity of our SiC microdisk resonator can be estimated by relating the amplitude of the thermomechanical motion to the RF power spectrum. For example, the spectral density of the thermomechanical motion at the resonance frequency is given by \cite{cleland2013foundations} 

\begin{equation}
S_{x,\text{th}}^{1/2}(\omega_m) = \sqrt{\frac{4k_B T Q_m}{\omega_m^3 m_{\text{eff}}}},
\end{equation}

\noindent where $k_B$ is the Boltzmann constant, $T$ is the temperature in Kelvin (300 K in our case), $\omega_m=2\pi f_m$, and $m_{\text{eff}}$ is the effective mass ($m_{\text{eff}} \approx 71$ pg for $4.3$-$\mu$m-radius SiC microdisk). Using this formula, the data shown in Fig.~\ref{Fig_R4p3}b translates to an exceptionally high transduction responsivity of $\sim143$ nV/fm and displacement sensitivity of $0.144$ $\text{fm/Hz}^{1/2}$. For convenience, we also plot the spectral density in the displacement domain on the right $y$-axis of Fig.~\ref{Fig_R4p3}c (details are referred to Section 1 in the Supplementary Information). 

In addition to the specific example ($4.3$-$\mu$m-radius SiC microdisk) discussed in Fig.~\ref{Fig_R4p3}, an array of SiC microdisks co-fabricated on the same chip are also characterized with the results summarized in Fig.~\ref{Fig_all}. Notably, we observe a consistent degradation of the mechanical quality factor with increased radius. This trend indicates that the mechanical quality factor is likely to be limited by pedestal-induced anchor loss, as the larger-sized microdisks are compromised by reduced undercutting ratios. This observation motivates future FEM simulations of energy dissipation pathways, specifically probing anchor-dependent losses and modal strain redistribution mechanisms amplified by the diminishing pedestal-to-radius ratio.

\subsection{Optomechanical self-oscillation and harmonic generation}
\begin{figure}[htbp]
\centering\includegraphics[width=0.9\linewidth]{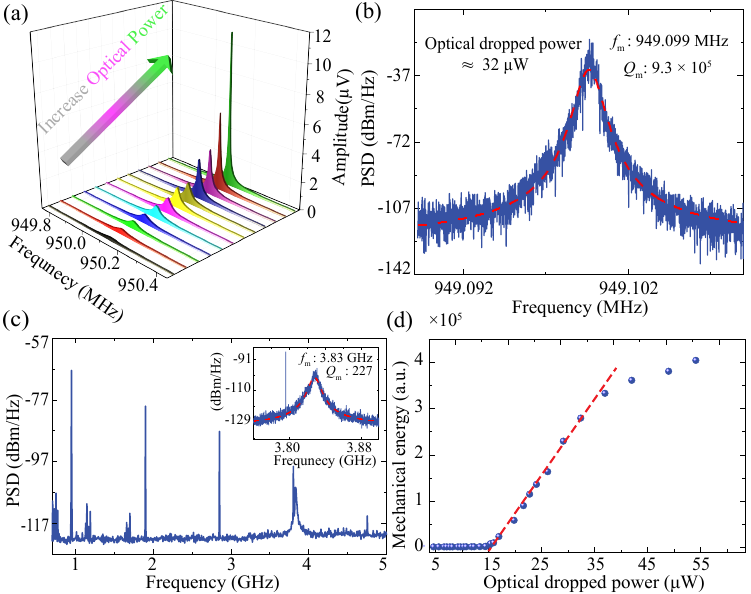} 
\caption{(a) Evolution of the photodetected RF spectrum near the resonant frequency of the fundamental radial breathing mode (RBM) as a function of the increased optical power. (b) Close-up view of the RF spectrum corresponding to a dropped power of 32 $\mu$W (resolution bandwidth of ESA set at 5 Hz). (c) Zoomed-out RF spectrum corresponding to the same optical dropped power in (b) (i.e., 32 $\mu$W) but with resolution bandwidth set at 5 kHz, revealing harmonics of the fundamental RBM as well as a secondary mechanical mode centered at $3.8$ GHz. The inset plots a closed-up view near the $3.8$-GHz mode (resolution bandwidth of the ESA set at 2 kHz), where the sharp peak represents the forth harmonics of the fundamental RBM. (d) Normalized mechanical energy as a function of the optical dropped power with the red dashed line representing a linear fit to the data above the threshold. }
\label{Fig_lasing}
\end{figure}
So far our mechanical characterization has been performed at low enough optical power levels to suppress the optical back-action, a necessary condition for the extraction of mechanical frequencies and quality factors in the linear regime. When operated at high enough optical power levels, however, the strong optomechanical back-action can surpass intrinsic mechanical damping and hence induce self-sustained regenerative optomechanical oscillations. We showcase such dynamic optomechanical interplay using the $4.3$-$\mu$m-radius SiC microdisk discussed in Fig.~\ref{Fig_R4p3} as an example. In Fig.~\ref{Fig_lasing}a, the measured RF spectrum of the fundamental RBM as a function of the optical powers is plotted. A close-up view of the mechanical spectrum at an approximate dropped power of $32\ \mu$W, shown in Fig.~\ref{Fig_lasing}b, reveals a significant reduction in the mechanical linewidth from the linear case of $\approx 50$ kHz to $\approx 1.0$ kHz, achieving an effective mechanical quality factor as high as $9.3\times10^5$. In addition, a full range of RF scan displays harmonic generation up to the fifth order for the fundamental RBM (Fig.~\ref{Fig_lasing}c). In the same spectrum, we also observe a secondary mechanical mode near the frequency of $3.8$ GHz. A zoomed-in display of this ultrahigh-frequency mode reveals a mechanical quality factor of 227 as shown in the inset of Fig.~\ref{Fig_lasing}c. While FEM simulations predict second- and third-order RBMs at $\sim 2.6$ GHz and $\sim 4.1$ GHz, respectively, both modes are absent in the measured RF spectrum. Instead, the observed $3.8$ GHz resonance is situated between these expected frequencies, suggesting a possible cause due to modal coupling between these high-order modes. Further investigation into the origin of this secondary mechanical mode is ongoing. 

Finally, the performance of our optomechanical oscillator can be more effectively quantified by plotting the mechanical energy as a function of the optical dropped power, as shown in Fig.~\ref{Fig_lasing}d. Here, the mechanical energy is obtained by integrating the transduced spectral density over the relevant frequency range, which is then normalized by the thermomechanical energy at room temperature in the absence of dynamic back-action. The resulting plot in Fig.~\ref{Fig_lasing}d exhibits a clear phonon-lasing behavior, featuring low mechanical energy at power levels below the threshold ($\approx 14\ \mu$W) and a dramatic increase with a large slope above the threshold. This threshold power is close to a theoretically predicted value of 19 $\mu$W (see Section 2 in the Supplementary Information), though we note the model used in earlier experiments has assumed a Lorentzian resonance which may require revision for optomechanical resonators showing mode splitting \cite{kippenberg_PRL_2005_analysis,jiang_OptExpress_2012_high, Mitchell_Optica_16_diamond}. When the optical power is large enough, the mechanical energy begins to saturate due to the optomechanically induced cavity frequency shifts approaching or exceeding the cavity linewidth.

\section{Discussions}
To benchmark the performance of our SiC optomechanical microdisk resonator, in particular in terms of the attained $f_m\cdot Q_m$ product, we compare our results to existing optomechanical and electromechanical resonators across common material platforms in Table 1. Note that this survey excludes designs based on the dissipation dilution technique, which allows mechanical resonators to achieve $f_m\cdot Q_m$ product above the intrinsic material (Akhiezer) limit \cite{maccabe_Science_2020_nano, engelsen_ultrahigh-quality-factor_2024}. As can be seen, the superior optical and mechanical properties of 4H-SiC have already elevated our $f_m\cdot Q_m$ product to be among the best reported numbers, with only a few experiments surpassing our result. Moreover, both the AlN \cite{han_APL_2015_10-ghz} and LN \cite{shen_APL_2020_high} experiments that reported higher $f_m\cdot Q_m$ products are for thin-film thickness modes that require non-optical excitation mechanisms. On the other hand, within the 4H-SiC realm, although a higher $f_m \cdot Q_m$ product has been reported in an earlier work \cite{hamelin_SciReports_2019_monocrystalline}, the reported mechanical resonator is substantially larger in dimension, requires external actuation, and can only maintain low dissipation in a vacuum environment. All these limiting factors can greatly restrict device miniaturization, integration, and scaling. In contrast, our devices are specifically designed for seamless integration with other photonic integrated circuit components and features a notably compact footprint that is comparable to the mechanical wavelength.

\begin{table}[htbp]
\centering
\caption{Survey of of reported optomechanical and electromechanical microresonators across common material platforms. For the $f_m\cdot Q_m$ column: $^{\ast}$ indicates results measured in vacuum; $^{\intercal}$ indicates results obtained in the cryogenic environment; and the rest without superscripts are conducted in ambient air at room temperature. $Q_0$ refers to the intrinsic optical quality factor.} 
\begin{adjustbox}{width=1\columnwidth,center}
\begin{tabular}{|c|c|c|c|c|c|c|}
\hline
\rowcolor[HTML]{9B9B9B} 
\textbf{Material} & \textbf{Reference} & \textbf{$f_m \cdot Q_m$ Product (THz)} & \textbf{$f_m$ (MHz)} & \textbf{$Q_m$} 
 ($\times10^4$) & \textbf{$m_{\text{eff}}$ (pg)} & \textbf{$Q_o$} ($\times 10^5$) \\ \hline

SiO$_2$ & M. ~Wang et. al.\cite{wang_PRX_2024_taming}   & 1.71         & 10167          & 0.02     & --                 & 22.7   \\ \hline

GaAs & N. C.~Carvalho et. al.\cite{c_APLPhotonics_2021_high}           & $12.6^{\intercal}$                         & 3150           & 0.4     & --                 & 0.3     \\ \hline

GaP & I. T.~Chen et. al.\cite{chen_NC_2023_optomechanical} & $8.19^{\ast\intercal}$                        & 2560           & 3200        & --         & 1.5            \\ \hline

InGaP & B.~Guha et. al.\cite{guha_OptExpress_2017_high}    & 0.7  & 481  & 0.15  & -- & 1.0 \\ \hline

Si  & W.~Jiang et. al.\cite{jiang_OptExpress_2012_high} & 4.3                         & 1294            & 0.33       & 5.7                & 3.5    \\ 
\hline

SiN & Y.~Liu et. al.\cite{liu_PRL_2013_electromagnetically} & $6.3^{\ast\intercal}$                         & 625             & 1       & 67                 & 10    \\ \hline

\multirow{2}{*}{AlN}& C.~Xiong et. al.\cite{xiong_APL_2012_integrated}  & 2.6                         & 1040             & 0.25       & 420                & 1.3        \\ 
\cline{2-7}
&X.~Han et. al.\cite{han_APL_2015_10-ghz}& 19.0                        & 10400             & 0.18     &   --              & 0.81        \\ \hline

\multirow{2}{*}{LiNbO$_3$} & W.~Jiang et. al.\cite{jiang_SciRep_2016_chip}  & 3.6                        & 1029           & 0.35       & 152                & 2.3      \\ \cline{2-7}
& M.~Shen et. al.\cite{shen_APL_2020_high} & $66^{\ast \intercal}$                         & 5200           & 1.25       & 0.69               & 21      \\ \hline

Diamond & M.~Mitchell et. al.\cite{Mitchell_Optica_16_diamond} & 19.0  & 2109  & 0.9   &40  & 0.7 \\ \hline

3C-SiC & X.~Lu et. al.\cite{lu_SciRep_2015_high} & 9.5                      & 1690             & 0.56       & 22                 & 0.4 \\
\hline

\multirow{3}{*}{4H-SiC} & L.~Sementilli et. al.\cite{sementilli_NanoLetters_2025_low} & $0.8^{\ast}$       & 0.053          & 1500     &  --       & \textbf{No}  \\ 
\cline{2-6}
& B.~Hamelin et. al. \cite{hamelin_SciReports_2019_monocrystalline}& $95^{\ast}$                      & 5.3            & 1800    &   --     & \textbf{Optical} \\ 
\cline{2-7}
& \textbf{This Work} & \textbf{18.2}                      & 951            & 1.92     & 71                 & 3.4  \\ \hline

\end{tabular}
\end{adjustbox}
\end{table}

% Please add the following required packages to your document preamble:
% \usepackage[table,xcdraw]{xcolor}
% Beamer presentation requires \usepackage{colortbl} instead of \usepackage[table,xcdraw]{xcolor}

\section{Conclusion}

In summary, we have demonstrated the first optomechanical resonators in the 4H-SiCOI platform based on ultracompact suspended 4H-SiC microdisks. These resonators, with radii varied from $4.3$ $\mu$m to $4.7$ $\mu$m, exhibit exceptional optical quality factors up to $\sim$1$\times10^6$, with their fundamental radial breathing mode featuring mechanical frequency around 950 MHz and mechanical quality factors up to $1.92\times10^4$. The corresponding $f_m \cdot Q_m$ product can be as high as $1.82 \times 10^{13}$ Hz, which is on par with the highest value reported among all whispering-gallery-mode optomechanical resonators measured in ambient. This advancement offers considerable promise for applications requiring high-precision sensing and metrology. In addition, the inherent material advantages of SiC render these devices well-suited for operations in extreme environments. Such unique combinations of high-performance sensing capabilities and robust material characteristics position SiC optomechanical systems as an ideal platform for addressing challenging applications where conventional technologies face significant limitations.

\newpage
\title{Supplementary Information}
\setcounter{equation}{0}
\setcounter{figure}{0}
\setcounter{table}{0}
\setcounter{section}{0}
\renewcommand{\thetable}{S\arabic{table}}
\renewcommand{\thefigure}{S\arabic{figure}}
\renewcommand{\theequation}{S\arabic{equation}}
\noindent 

\section{Displacement sensitivity calibration}
The total electronic-domain noise spectral density is $S_{v,\text{total}}^{1/2} = \left( S_{v,\text{th}} + S_{v,\text{sys}} \right)^{1/2}$, where $S_{v,\text{th}}^{1/2}$ and $S_{v,\text{th}}^{1/2}$ represent the contributions from thermomechanical motion and measurement system, respectively. 
$S_{v,\text{sys}}^{1/2}$ sets the off-resonance background ($S_{v,\text{total}}^{1/2} \approx S_{v,\text{sys}}^{1/2}$ when $\omega \ne \omega_m $), and $S_{v,\text{th}}^{1/2}$ is related to $S_{x,\text{th}}^{1/2}$ through the `displacement-to-voltage' responsivity $\mathfrak{R} \equiv S_{v,\text{th}}^{1/2}/S_{x,\text{th}}^{1/2}$. The displacement sensitivity of the measurement system is defined as $S_{x,\text{sys}}^{1/2}=S_{v,\text{sys}}^{1/2}/\mathfrak{R}$ and can be estimated as\cite{wang_NatCommun_2014_spatial}:

\begin{align}
S_{x,\text{sys}}^{1/2} &= \frac{1}{\mathfrak{R}} S_{v,\text{total}}^{1/2} (\omega \ne \omega_m) = 
\frac{S_{x,\text{th}}^{1/2} (\omega_m)}{S_{v,\text{th}}^{1/2} (\omega_m)} 
S_{v,\text{total}}^{1/2} (\omega \ne \omega_m), \notag \\
&= \frac{S_{x,\text{th}}^{1/2} (\omega_m)}
{\sqrt{S_{v,\text{total}} (\omega_m) - S_{v,\text{sys}} (\omega_m)}}
S_{v,\text{total}}^{1/2} (\omega \ne \omega_m). \tag{S1}
\end{align}

\noindent Assuming $S_{v,\text{sys}}^{1/2}$ is constant across the frequency range, we use $S_{v,\text{sys}}^{1/2}=$13 $\mathrm{n V/Hz^{1/2}}$ and $S_{v,\text{tot}}^{1/2}=$24 $\mathrm{n V/Hz^{1/2}}$ determined from Fig.~4b, and obtain $S_{x,\text{sys}}^{1/2}=$ 0.093 $\mathrm{fm/Hz^{1/2}}$. The responsivity can thus be calculated as $\mathfrak{R} \equiv S_{v,\text{th}}^{1/2}/S_{x,\text{th}}^{1/2}=143$ nV/fm, which allows us to plot displacement domain noise spectral density on the right y-axis of Fig.~4c through the relationship $S_{x,\text{th}}^{1/2}=S_{v,\text{th}}^{1/2}/\mathfrak{R}$.

\section{Analysis of power threshold of optomechanical oscillation}

The threshold power corresponding to optomechanical oscillation, assuming a Lorentizan resonance, is given by \cite{kippenberg_PRL_2005_analysis,jiang_OptExpress_2012_high}:

\begin{equation}
P_d = \frac{m_{\text{eff}} \omega_0}{2 g_{\text{OM}}^2} 
\frac{\Gamma_0 \Gamma_m}{\Gamma_t \Delta} 
\left[ (\Delta - \Omega_m)^2 + \left(\frac{\Gamma_t}{2}\right)^2 \right] 
\left[ (\Delta + \Omega_m)^2 + \left(\frac{\Gamma_t}{2}\right)^2 \right], \tag{S2}
\end{equation}

\noindent where $P_d$ denotes the threshold power dropped into the microdisk, with the definition and numerical values of other parameters explained in Table S1. Based on Eq.~S2, we obtain a theoretical threshold power of $\approx$ 19 $\mu$W, which is slightly above the experimental data of $14$ $\mu$W (see Fig.~6d in the main text). We believe the discrepancy suggests that Eq.~S2 may require revision for resonances showing mode splitting, in particular when the splitting frequency is on the same order as the mechanical frequency. In addition, it is likely important to include other contributions such as the photoelastic effect, which could play an important role determining the overall optomechanical coupling \cite{Balram_Optica_2014_GaAs}. A more comprehensive model incorporating these factors is planned for future work.  

\renewcommand{\thetable}{S\arabic{table}}
\begin{table}[htbp]
\captionsetup{justification=centering}
\centering
\caption{Summary of key parameters used in the power threshold calculation for the optomechanical oscillator}
\begin{adjustbox}{width=0.7\columnwidth,center}
\begin{tabular}{c|c|c}
\hline
\textbf{Parameters}              & \textbf{Description}                & \textbf{Values}   \\ \hline
$m_{\text{eff}}$                 & Effective mass                      & 71 pg             \\
$\omega_0 / 2\pi$                & Optical frequency                   & 188.56 THz        \\
$g_{\text{OM}}/ 2\pi$                   & Optomechanical coupling coefficient & 44 GHz/nm         \\
$\Gamma_0$                       & Intrinsic photon decay rate         & 3.48 ns$^{-1}$    \\
$\Gamma_m$                       & Mechanical energy decay rate        & 0.31 $\mu\text{s}^{-1}$ \\
$\Gamma_t$                       & Loaded photon decay rate            & 3.70 ns$^{-1}$    \\
$\Delta$                         & Laser-cavity detuning               & 760 MHz           \\
$\Omega_m / 2\pi$                & Mechanical frequency                & 950.14 MHz        \\ \hline 
\end{tabular}
\end{adjustbox}
\end{table}

\begin{backmatter}
\bmsection{Funding}
The CMU team was supported by NSF (2131162, 2240420). Y.~Liu and P.~Feng are thankful to the partial support from NSF IUCRC MIST Center (Grant EEC-1939009), NSF Central Florida Semiconductor Innovation Engine (Grant ITE-2315320), and DARPA DSO OpTIm Program (Grant HR00112320028).
\bmsection{Acknowledgments}
The authors would like to acknowledge helpful discussions with Prof.~Gianluca Piazza, as well as contributions from Dr.~Ruixuan Wang and Sam Gou in the early stages of the experiment. The authors also acknowledge the use of Bertucci Nanotechnology Laboratory at Carnegie Mellon University supported by grant BNL-78657879 and the Materials Characterization Facility supported by grant MCF-677785. 

\bmsection{Disclosures}  The authors declare no conflicts of interest.

\bmsection{Data Availability} Data underlying the results presented in this paper are not publicly available at this time but may be obtained from the authors upon reasonable request.

\end{backmatter}

\bibliography{MyRef}

\begin{thebibliography}{10}
\newcommand{\enquote}[1]{``#1''}

\bibitem{aspelmeyerRMP2014cavity}
M.~Aspelmeyer, T.~J. Kippenberg, and F.~Marquardt, \enquote{Cavity optomechanics,} {\protect\JournalTitle{Reviews of Modern Physics}} \textbf{86}, 1391--1452 (2014).

\bibitem{hossein_PRA2006_char}
M.~Hossein-Zadeh, H.~Rokhsari, A.~Hajimiri, and K.~J. Vahala, \enquote{Characterization of a radiation-pressure-driven micromechanical oscillator,} {\protect\JournalTitle{Physical Review A}} \textbf{74}, 023813 (2006).

\bibitem{tallur_OptExpress_2011_monolithic}
S.~Tallur, S.~Sridaran, and S.~A. Bhave, \enquote{A monolithic radiation-pressure driven, low phase noise silicon nitride opto-mechanical oscillator,} {\protect\JournalTitle{Optics Express}} \textbf{19}, 24522--24529 (2011).

\bibitem{luan_SciRep_2014_integrated}
X.~Luan, Y.~Huang, Y.~Li, J.~F. McMillan, J.~Zheng, S.-W. Huang, P.-C. Hsieh, T.~Gu, D.~Wang, A.~Hati \emph{et~al.}, \enquote{An integrated low phase noise radiation-pressure-driven optomechanical oscillator chipset,} {\protect\JournalTitle{Scientific Reports}} \textbf{4}, 6842 (2014).

\bibitem{safavi_Nature_2013_squeezed}
A.~H. Safavi-Naeini, S.~Gr{\"o}blacher, J.~T. Hill, J.~Chan, M.~Aspelmeyer, and O.~Painter, \enquote{Squeezed light from a silicon micromechanical resonator,} {\protect\JournalTitle{Nature}} \textbf{500}, 185--189 (2013).

\bibitem{bochmann_NatPhys_2013_nanomechanical}
J.~Bochmann, A.~Vainsencher, D.~D. Awschalom, and A.~N. Cleland, \enquote{Nanomechanical coupling between microwave and optical photons,} {\protect\JournalTitle{Nature Physics}} \textbf{9}, 712--716 (2013).

\bibitem{balram_NatPhotonics_2016_coherent}
K.~C. Balram, M.~I. Davan{\c{c}}o, J.~D. Song, and K.~Srinivasan, \enquote{Coherent coupling between radiofrequency, optical and acoustic waves in piezo-optomechanical circuits,} {\protect\JournalTitle{Nature Photonics}} \textbf{10}, 346--352 (2016).

\bibitem{wu_NatNanotechnol_2017_nanocavity}
M.~Wu, N.~L.-Y. Wu, T.~Firdous, F.~Fani~Sani, J.~E. Losby, M.~R. Freeman, and P.~E. Barclay, \enquote{Nanocavity optomechanical torque magnetometry and radiofrequency susceptometry,} {\protect\JournalTitle{Nature Nanotechnology}} \textbf{12}, 127--131 (2017).

\bibitem{li_Nanophotonics_2021_cavity}
B.-B. Li, L.~Ou, Y.~Lei, and Y.-C. Liu, \enquote{Cavity optomechanical sensing,} {\protect\JournalTitle{Nanophotonics}} \textbf{10}, 2799--2832 (2021).

\bibitem{sun_APL_2012_high}
X.~Sun, X.~Zhang, and H.~X. Tang, \enquote{High-{Q} silicon optomechanical microdisk resonators at gigahertz frequencies,} {\protect\JournalTitle{Applied Physics Letters}} \textbf{100} (2012).

\bibitem{schwab_MN_2022_very}
L.~Schwab, P.~Allain, N.~Mauran, X.~Dollat, L.~Mazenq, D.~Lagrange, M.~G{\'e}ly, S.~Hentz, G.~Jourdan, I.~Favero \emph{et~al.}, \enquote{Very-high-frequency probes for atomic force microscopy with silicon optomechanics,} {\protect\JournalTitle{Microsystems \& Nanoengineering}} \textbf{8}, 32 (2022).

\bibitem{sonar_Optica_2025_high}
S.~Sonar, U.~Hatipoglu, S.~Meesala, D.~P. Lake, H.~Ren, and O.~Painter, \enquote{High-efficiency low-noise optomechanical crystal photon-phonon transducers,} {\protect\JournalTitle{Optica}} \textbf{12}, 99--104 (2025).

\bibitem{wu_SciAdvances_2023_chip}
N.~Wu, K.~Cui, Q.~Xu, X.~Feng, F.~Liu, W.~Zhang, and Y.~Huang, \enquote{On-chip mechanical exceptional points based on an optomechanical zipper cavity,} {\protect\JournalTitle{Science Advances}} \textbf{9}, eabp8892 (2023).

\bibitem{kippenberg_PRL_2005_analysis}
T.~Kippenberg, H.~Rokhsari, T.~Carmon, A.~Scherer, and K.~Vahala, \enquote{Analysis of radiation-pressure induced mechanical oscillation of an optical microcavity,} {\protect\JournalTitle{Physical Review Letters}} \textbf{95}, 033901 (2005).

\bibitem{jiang_OptExpress_2015_chip}
X.~Jiang, M.~Wang, M.~C. Kuzyk, T.~Oo, G.-L. Long, and H.~Wang, \enquote{Chip-based silica microspheres for cavity optomechanics,} {\protect\JournalTitle{Optics Express}} \textbf{23}, 27260--27265 (2015).

\bibitem{wang_PRX_2024_taming}
M.~Wang, Z.-G. Hu, C.~Lao, Y.~Wang, X.~Jin, X.~Zhou, Y.~Lei, Z.~Wang, W.~Liu, Q.-F. Yang \emph{et~al.}, \enquote{Taming brillouin optomechanics using supermode microresonators,} {\protect\JournalTitle{Physical Review X}} \textbf{14}, 011056 (2024).

\bibitem{liu_PRL_2013_electromagnetically}
Y.~Liu, M.~Davan{\c{c}}o, V.~Aksyuk, and K.~Srinivasan, \enquote{Electromagnetically induced transparency and wideband wavelength conversion in silicon nitride microdisk optomechanical resonators,} {\protect\JournalTitle{Physical Review Letters}} \textbf{110}, 223603 (2013).

\bibitem{grutter_APLPhotonics_2018_invited}
K.~E. Grutter, M.~I. Davan{\c{c}}o, K.~C. Balram, and K.~Srinivasan, \enquote{Invited article: Tuning and stabilization of optomechanical crystal cavities through nems integration,} {\protect\JournalTitle{APL Photonics}} \textbf{3} (2018).

\bibitem{han_APL_2015_10-ghz}
X.~Han, K.~Y. Fong, and H.~X. Tang, \enquote{A 10-{GHz} film-thickness-mode cavity optomechanical resonator,} {\protect\JournalTitle{Applied Physics Letters}} \textbf{106} (2015).

\bibitem{sohn_NatPhotonics_2018_time}
D.~B. Sohn, S.~Kim, and G.~Bahl, \enquote{Time-reversal symmetry breaking with acoustic pumping of nanophotonic circuits,} {\protect\JournalTitle{Nature Photonics}} \textbf{12}, 91--97 (2018).

\bibitem{Mitchell_Optica_16_diamond}
M.~Mitchell, B.~Khanaliloo, D.~P. Lake, T.~Masuda, J.~P. Hadden, and P.~E. Barclay, \enquote{Single-crystal diamond low-dissipation cavity optomechanics,} {\protect\JournalTitle{Optica}} \textbf{3}, 963--970 (2016).

\bibitem{burek_Optica_2016_diamond}
M.~J. Burek, J.~D. Cohen, S.~M. Meenehan, N.~El-Sawah, C.~Chia, T.~Ruelle, S.~Meesala, J.~Rochman, H.~A. Atikian, M.~Markham \emph{et~al.}, \enquote{Diamond optomechanical crystals,} {\protect\JournalTitle{Optica}} \textbf{3}, 1404--1411 (2016).

\bibitem{schneider_Optica_2019_optomechanics}
K.~Schneider, Y.~Baumgartner, S.~H{\"o}nl, P.~Welter, H.~Hahn, D.~J. Wilson, L.~Czornomaz, and P.~Seidler, \enquote{Optomechanics with one-dimensional gallium phosphide photonic crystal cavities,} {\protect\JournalTitle{Optica}} \textbf{6}, 577--584 (2019).

\bibitem{chen_NC_2023_optomechanical}
I.-T. Chen, B.~Li, S.~Lee, S.~Chakravarthi, K.-M. Fu, and M.~Li, \enquote{Optomechanical ring resonator for efficient microwave-optical frequency conversion,} {\protect\JournalTitle{Nature Communications}} \textbf{14}, 7594 (2023).

\bibitem{jiang_SciRep_2016_chip}
W.~C. Jiang and Q.~Lin, \enquote{Chip-scale cavity optomechanics in lithium niobate,} {\protect\JournalTitle{Scientific Reports}} \textbf{6}, 36920 (2016).

\bibitem{shen_APL_2020_high}
M.~Shen, J.~Xie, C.-L. Zou, Y.~Xu, W.~Fu, and H.~X. Tang, \enquote{High frequency lithium niobate film-thickness-mode optomechanical resonator,} {\protect\JournalTitle{Applied Physics Letters}} \textbf{117} (2020).

\bibitem{jiang_Optica_2019_lithium}
W.~Jiang, R.~N. Patel, F.~M. Mayor, T.~P. McKenna, P.~Arrangoiz-Arriola, C.~J. Sarabalis, J.~D. Witmer, R.~Van~Laer, and A.~H. Safavi-Naeini, \enquote{Lithium niobate piezo-optomechanical crystals,} {\protect\JournalTitle{Optica}} \textbf{6}, 845--853 (2019).

\bibitem{falk_NC_2013_polytype}
A.~L. Falk, B.~B. Buckley, G.~Calusine, W.~F. Koehl, V.~V. Dobrovitski, A.~Politi, C.~A. Zorman, P.~X.-L. Feng, and D.~D. Awschalom, \enquote{Polytype control of spin qubits in silicon carbide,} {\protect\JournalTitle{Nature Communications}} \textbf{4}, 1819 (2013).

\bibitem{lu_APL_2020_silicon}
X.~Lu, J.~Y. Lee, and Q.~Lin, \enquote{Silicon carbide zipper photonic crystal optomechanical cavities,} {\protect\JournalTitle{Applied physics letters}} \textbf{116} (2020).

\bibitem{sementilli_NanoLetters_2025_low}
L.~Sementilli, D.~M. Lukin, H.~Lee, J.~Yang, E.~Romero, J.~Vučković, and W.~P. Bowen, \enquote{Low-dissipation nanomechanical devices from monocrystalline silicon carbide,} {\protect\JournalTitle{Nano Letters}} \textbf{25}, 6069--6075 (2025).

\bibitem{liu_CommuniEngi_2024_4h}
Z.~Liu, Y.~Long, C.~Wehner, H.~Wen, and F.~Ayazi, \enquote{4{H} silicon carbide bulk acoustic wave gyroscope with ultra-high {Q}-factor for on-chip inertial navigation,} {\protect\JournalTitle{Communications Engineering}} \textbf{3} (2024).

\bibitem{chandorkar2008limits}
S.~Chandorkar, M.~Agarwal, R.~Melamud, R.~Candler, K.~Goodson, and T.~Kenny, \enquote{Limits of quality factor in bulk-mode micromechanical resonators,} in \emph{2008 IEEE 21st international conference on micro electro mechanical systems,}  (IEEE, 2008), pp. 74--77.

\bibitem{ayazi2011energy}
F.~Ayazi, L.~Sorenson, and R.~Tabrizian, \enquote{Energy dissipation in micromechanical resonators,} in \emph{Micro-and Nanotechnology Sensors, Systems, and Applications III,}  vol. 8031 (SPIE, 2011), pp. 371--383.

\bibitem{ghaffari_SciRep_2013_quantum}
S.~Ghaffari, S.~A. Chandorkar, S.~Wang, E.~J. Ng, C.~H. Ahn, V.~Hong, Y.~Yang, and T.~W. Kenny, \enquote{Quantum limit of quality factor in silicon micro and nano mechanical resonators,} {\protect\JournalTitle{Scientific Reports}} \textbf{3}, 3244 (2013).

\bibitem{lu_SciRep_2015_high}
X.~Lu, J.~Y. Lee, and Q.~Lin, \enquote{High-frequency and high-quality silicon carbide optomechanical microresonators,} {\protect\JournalTitle{Scientific Reports}} \textbf{5}, 17005 (2015).

\bibitem{Li_4HSiC_comb}
L.~Cai, J.~Li, R.~Wang, and Q.~Li, \enquote{Octave-spanning microcomb generation in {4H}-silicon-carbide-on-insulator photonics platform,} {\protect\JournalTitle{Photonics Research}} \textbf{10}, 870--876 (2022).

\bibitem{YiXu_mode_splitting}
X.~Yi, Y.-F. Xiao, Y.-C. Liu, B.-B. Li, Y.-L. Chen, Y.~Li, and Q.~Gong, \enquote{Multiple-{Rayleigh}-scatterer-induced mode splitting in a high-\${Q}\$ whispering-gallery-mode microresonator,} {\protect\JournalTitle{Physical Review A}} \textbf{83}, 023803 (2011).

\bibitem{Li_azimuth}
Q.~Li, A.~A. Eftekhar, Z.~Xia, and A.~Adibi, \enquote{Azimuthal-order variations of surface-roughness-induced mode splitting and scattering loss in high-{Q} microdisk resonators,} {\protect\JournalTitle{Optics Letters}} \textbf{37}, 1586--1588 (2012).

\bibitem{jia_JMEMS_2019_very}
H.~Jia and P.~X.-L. Feng, \enquote{Very high-frequency silicon carbide microdisk resonators with multimode responses in water for particle sensing,} {\protect\JournalTitle{Journal of Microelectromechanical Systems}} \textbf{28}, 941--953 (2019).

\bibitem{cleland2013foundations}
A.~N. Cleland, \emph{Foundations of nanomechanics: from solid-state theory to device applications} (Springer Science \& Business Media, 2013).

\bibitem{jiang_OptExpress_2012_high}
W.~C. Jiang, X.~Lu, J.~Zhang, and Q.~Lin, \enquote{High-frequency silicon optomechanical oscillator with an ultralow threshold,} {\protect\JournalTitle{Optics Express}} \textbf{20}, 15991--15996 (2012).

\bibitem{maccabe_Science_2020_nano}
G.~S. MacCabe, H.~Ren, J.~Luo, J.~D. Cohen, H.~Zhou, A.~Sipahigil, M.~Mirhosseini, and O.~Painter, \enquote{Nano-acoustic resonator with ultralong phonon lifetime,} {\protect\JournalTitle{Science}} \textbf{370}, 840--843 (2020).

\bibitem{engelsen_ultrahigh-quality-factor_2024}
N.~J. Engelsen, A.~Beccari, and T.~J. Kippenberg, \enquote{Ultrahigh-quality-factor micro- and nanomechanical resonators using dissipation dilution,} {\protect\JournalTitle{Nature Nanotechnology}} \textbf{19}, 725--737 (2024).

\bibitem{hamelin_SciReports_2019_monocrystalline}
B.~Hamelin, J.~Yang, A.~Daruwalla, H.~Wen, and F.~Ayazi, \enquote{Monocrystalline silicon carbide disk resonators on phononic crystals with ultra-low dissipation bulk acoustic wave modes,} {\protect\JournalTitle{Scientific Reports}} \textbf{9} (2019).

\bibitem{c_APLPhotonics_2021_high}
N.~C~Carvalho, R.~Benevides, M.~M{\'e}nard, G.~S~Wiederhecker, N.~C~Frateschi, and T.~Mayer~Alegre, \enquote{High-frequency {GaAs} optomechanical bullseye resonator,} {\protect\JournalTitle{APL photonics}} \textbf{6} (2021).

\bibitem{guha_OptExpress_2017_high}
B.~Guha, S.~Mariani, A.~Lemaitre, S.~Combrié, G.~Leo, and I.~Favero, \enquote{High frequency optomechanical disk resonators in {III}–{V} ternary semiconductors,} {\protect\JournalTitle{Optics Express}} \textbf{25}, 24639–24639 (2017).

\bibitem{xiong_APL_2012_integrated}
C.~Xiong, X.~Sun, K.~Y. Fong, and H.~X. Tang, \enquote{Integrated high frequency aluminum nitride optomechanical resonators,} {\protect\JournalTitle{Applied Physics Letters}} \textbf{100} (2012).

\bibitem{wang_NatCommun_2014_spatial}
Z.~Wang, J.~Lee, and P.~X.-L. Feng, \enquote{Spatial mapping of multimode brownian motions in high-frequency silicon carbide microdisk resonators,} {\protect\JournalTitle{Nature Communications}} \textbf{5}, 5158 (2014).

\bibitem{Balram_Optica_2014_GaAs}
K.~C. Balram, M.~Davan\c{c}o, J.~Y. Lim, J.~D. Song, and K.~Srinivasan, \enquote{Moving boundary and photoelastic coupling in {GaAs} optomechanical resonators,} {\protect\JournalTitle{Optica}} \textbf{1}, 414--420 (2014).

\end{thebibliography}
\end{document}